\documentclass[fleqn,usenatbib]{mnras}

\usepackage{newtxtext,newtxmath}

\usepackage[T1]{fontenc}

\DeclareRobustCommand{\VAN}[3]{#2}
\let\VANthebibliography\thebibliography
\def\thebibliography{\DeclareRobustCommand{\VAN}[3]{##3}\VANthebibliography}

\usepackage{graphicx}	
\usepackage{amsmath}	
\usepackage{rotating}
\usepackage{xcolor}
\usepackage{mathtools}

\title[Partially Cloudy]{Mapping the Surface of Partially Cloudy Exoplanets is Hard}

\author[Teinturier et al.]{
Lucas Teinturier $^{1,2}$\thanks{E-mail: lucas.teinturier@mail.mcgill.ca},
Nicholas Vieira$^{1}$,
Elisa Jacquet$^{1,3}$,
Juliette Geoffrion$^{1,4}$,
\newauthor{Youssef Bestavros$^{1}$, Dylan Keating$^{1}$}
 and Nicolas B. Cowan$^{1,5}$
\\
$^{1}$ Department of Physics, McGill University, 3600 rue University, Montréal, QC Canada H3A 2T8\\
$^{2}$ Ecole CentraleSupelec, 3 Rue Joliot-Curie 91192, Gif-Sur-Yvette, France \\
$^{3}$ Physics Department, Imperial College, London SW7 2BZ \\
$^{4}$ Département de physique, Université de Montréal, Montréal (Québec), H3C 3J7, Canada\\
$^{5}$ Department of Earth \& Planetary Sciences, McGill University, 3450 rue University, Montréal, QC Canada, H3A 0E8
}

\date{Accepted XXX. Received YYY; in original form ZZZ}

\pubyear{2020}

\begin{document}
\label{firstpage}
\pagerange{\pageref{firstpage}--\pageref{lastpage}}
\maketitle

\begin{abstract}
Reflected light photometry of terrestrial exoplanets could reveal the presence of oceans and continents, hence placing direct constraints on the current and long-term habitability of these worlds. Inferring the albedo map of a planet from its observed light curve is challenging because different maps may yield indistinguishable light curves. This degeneracy is aggravated by changing clouds.  It has previously been suggested that disk-integrated photometry spanning multiple days could be combined to obtain a cloud-free surface map of an exoplanet.  We demonstrate this technique as part of a Bayesian retrieval by simultaneously fitting for the fixed surface map of a planet and the time-variable overlying clouds. We test this approach on synthetic data then apply it to real disk-integrated observations of the Earth. We find that eight days of continuous synthetic observations are sufficient to reconstruct a faithful low resolution surface albedo map, without needing to make assumptions about cloud physics. For lightcurves with negligible photometric uncertainties, the minimal top-of-atmosphere albedo at a location is a good estimate of its surface albedo. When applied to observations from the Earth Polychromating Imaging Camera aboard the DSCOVR spacecraft, our approach removes only a small fraction of clouds. We attribute this difficulty to the full-phase geometry of observations combined with the short correlation length for Earth clouds. For exoplanets with Earth-like climatology, it may be hard to do much better than a cloud-averaged map. We surmise that cloud removal will be most successful for exoplanets imaged near quarter phase that harbour large cloud systems.
\end{abstract}

\begin{keywords}
planets and satellites: atmospheres - planets and satellites: surfaces - planets and satellites: terrestrial planets - method: numerical
\end{keywords}



\section{Introduction}

Surface maps of terrestrial exoplanets could constrain their current and long-term habitability \citep{abbot_indication_2012,Robinson_2018}. Indeed, surface  albedo maps could identify planetary features such as oceans, continents, or even vegetation (\citealt{cowan_alien_2009,cowan_rotational_2011,fujii_colors_2010,fujii_colors_2011,2011ESS.....2.2903F,fujii_rotational_2017,kawahara_global_2010,kawahara_fujii_2012,lustig-yaeger_detecting_2018}). For a review of exo-cartography, see \cite{co02010y}.  \par
Exo-cartography of an Earth-like planet orbiting a Sun-like star will require direct-imaging. The \citet{NAP25187}'s Exoplanet Science Strategy report states that even though direct imaging of mature planet in thermal emission is easier than in reflected light in terms of flux ratio, the technical challenges associated with imaging at longer wavelengths are greater. Reflected light is believed to be a more straightforward path towards directly imaging Earth twins and mapping their surface. In particular, a space-based mission such as LUVOIR \citep{2019LUVOIR} or HABEX \citep{gaudi2020habitable}, equiped with a coronograph or a starshade would be a great asset to characterize Earth-size planets in reflected light. \citet{NAP26141}'s Pathways to Discovery in Astronomy and Astrophysics for the 2020's recommends a large infrared-optical-ultraviolet space telescope with high-contrast imaging and spectroscopy. This new space observatory could provide the necessary reflected light to perform exo-cartography. \par 
Mapping the atmosphere or surface of an exoplanet based on its time-varying reflectivity is an under-constrained inverse problem: different maps can produce the same light curve \citep{10.1093/mnras/stt1191}. These degeneracies are  made worse by clouds that change from one planetary rotation to another (\citealt{cowan_alien_2009,2011ESS.....2.2903F}). Previous research has either neglected clouds altogether or solved for a time-averaged top-of-atmosphere (TOA) albedo map (\citealt{cowan_alien_2009,cowan_rotational_2011,farr_exocartographer_2018,fujii_colors_2011,kawahara_global_2010}). \cite{kawahara_bayesian_2020} instead accounted for the changing cloud cover by explicitly modelling the time-variable TOA albedo maps, but did not attempt to reconstruct a cloud-free surface map. \citet{2019arXivLuger} attempted to fit for a time-varying cloud map, using serendipitous \textit{TESS} photometry of Earth. However, they do not estimate uncertainties on their retrieved map, as they simply compute the maximal likelihood solution with respect to their constraints. They detected time variable clouds, but concluded that future work was warranted. \par

We set out to map the surface of an unresolved planet using reflected light photometry despite time-variable clouds. Over the course of days, clouds move and disperse. One expects that the minimum top-of-atmosphere albedo of some location will correspond to the least cloudy epoch---and hence most closely resemble the surface albedo. To provide more robust uncertainty estimates on the surface albedo, we will simultaneously fit for the invariable surface albedo and the changing cloud cover of the planet, with the goal of retrieving a faithful surface albedo map.

In Section \ref{cloud model}, we present our time-variable cloud model and the generation of synthetic lightcurves. In Section \ref{Section : remov synth data}, we demonstrate the removal of clouds from synthetic data. In Section \ref{EPIC} we test our scheme on data from the Deep Space Climate Observatory's Earth Polychromatic Imaging Camera. We discuss our results in Section \ref{Discuss} and we conclude in Section \ref{conclu}.

\section{Producing Synthetic Lightcurves } \label{cloud model}
In this section, we present the time-variable cloud model we will use to generate synthetic lightcurves, as well as in all of our retrievals

\subsection{Forward Model}
We longitudinally split the surface of a planet into equal width slices, each with uniform cloud and surface albedo. While the effect of clouds on top-of-atmosphere albedo is complex, the net shortwave effect is usually to increase the TOA albedo because the albedo of clouds is greater than that of most plausible surfaces. As a proof of principle, we therefore adopt the very simplest model to combine the surface and the cloud albedo: 

\begin{equation} \label{effAlb} 
    A_{{\rm TOA},i,j} = A_{{\rm s},i}+A_{c,i,j}
\end{equation}
where the subscript $i$ denotes the slice and the subscript $j$ the day, $\rm A_{s,i}$ being the surface albedo of slice $\rm i$, and $\rm A_{c,i,j}$ the increase of albedo due to clouds in slice $i$, on day $j$. 

\subsection{Viewing Geometry and Light Curves}

We quantify the planet's reflectivity by its apparent albedo \citep{2003JGRD..108.4709Q,10.1093/mnras/stt1191}:
\begin{equation}\label{forwardmodel}
    \centering
    A^*(t) =\frac{ \oint K(\Omega,t)A_{\rm TOA }(\Omega)d\Omega}{ \oint K(\Omega,t) d\Omega}
\end{equation}
where $\Omega$ specifies the co-latitude and longitude $(\theta,\phi)$, the convolution kernel $K(\Omega,t)$ is a function of location and time, and the differential solid angle is $d\Omega = \sin\theta d\theta d\phi$. If we presume Lambertian reflection, then the kernel is simply the normalised product of visibility $V$ and illumination $I$: $\frac{1}{\pi}V(\Omega,t)I(\Omega,t)$.\par 
The visibility and illumination functions are
\begin{align} \small
    V = \max[ \sin\theta \sin\theta_o \cos(\phi - \phi_o) + \cos\theta \cos\theta_o , 0] \label{visu}\\
    I = \max[ \sin\theta \sin\theta_s \cos(\phi - \phi_s) + \cos\theta \cos\theta_s , 0]\label{illu} ,
\end{align}
 where the subscript $o$ denotes the sub-observer point, and the subscript $s$ the sub-stellar point. 
 
 To streamline comparison to Earth data, we adopt the viewing geometry of the Deep Space Climate Observatory, located at the first Earth--Sun Lagrange point (L$_1$). For an observer at L$_1$, the sub-stellar and sub-observer points are coincident: $\theta_s = \theta_o$ and $\phi_s = \phi_o$. Since we will only use data spanning a few days, we assume that both sub-stellar and sub-observer co-latitudes are equal to $\frac{\pi}{2}$, which is exactly correct at the equinoxes. The sub-observer longitude is given by $\phi_o$ = $ \phi_o(0) - \omega t$, where $\omega$ is the rotational angular speed of the planet.  One should also note that $\phi = 0$ is defined at the prime meridian and longitude increases eastwards. \par

 In this work we only consider longitudinal variations in albedo but allow for a time-variable albedo map so we replace $A_{\rm TOA}(\Omega)$ with $A_{\rm TOA}$($\phi$,t). Therefore, the apparent albedo of the planet is
\begin{equation} \small \label{equation sur Astar}
    A^*(t) =\frac { \frac{1}{\pi} \int_{IV} A_{\rm TOA}(\phi,t) \sin^3(\theta) \cos^2(\phi-\phi_o)d \theta d \phi}{\frac{1}{\pi} \int_{IV} \sin^3(\theta) \cos^2(\phi-\phi_o)d \theta d\phi} ,
\end{equation}
where the integral spans only the illuminated and visible lune, $IV$. For an observer at L$_1$, this lune is simply the dayside hemisphere.

It is convenient to break down Eq. (\ref{equation sur Astar}) into contributions from each slice. After integrating over latitude, a single fully visible longitudinal slice $i$ spanning longitudes $[\phi_i, \phi_{i+1}]$ contributes to the instantaneous disk integrated albedo as
\begin{equation}
    A^*_i(t) =\frac{2}{\pi}A_{\rm TOA,i} \int_{\phi_i}^{\phi_{i+1}} \cos^2\left(\phi - \phi_o(t)\right) d \phi.
\end{equation}
 
 The planet's apparent albedo $A^*(t)$ is the sum of the contributions from  all $m$ visible slices : 
\begin{equation}\label{forward mslices} \small
    A^*(t) =\frac{2}{\pi}  \sum_{i = 1}^{m} A_{ \rm TOA,i} \int_{\max(T_W,\phi_i)}^{\min(T_E, \phi_{i+1})} \cos^2(\phi - \phi_o) d\phi ,
\end{equation}
where T$_E$ and T$_W$ are the East and West terminators, respectively. These limits of integration ensure that regions on the far side of the planet cannot contribute reflected flux.

Integrating Eq. (\ref{forward mslices}) yields
\begin{equation}\label{forward_fin} \small
    A^*(t) =\frac{1}{\pi}  \sum_{i = 1}^{m} A_{ \rm TOA,i}\Bigg[ \phi +\frac{\sin(2 \phi - 2 \phi_o)}{2}\Bigg]_{\max(T_W,\phi_i)}^{\min(T_E,\phi_{i+1})}
\end{equation}
for $m$ slices between the West and East terminators. By combining Eqs. (\ref{effAlb}) and (\ref{forward_fin}), we can now rapidly compute the light curve of a planet given its surface albedo map and time-variable cloud map. We provide a derivation of the forward model for arbitrary orbital phase in Appendix \ref{appendix}.

\subsection{Synthetic lightcurves}
\begin{figure*}

    \includegraphics[scale = 0.64]{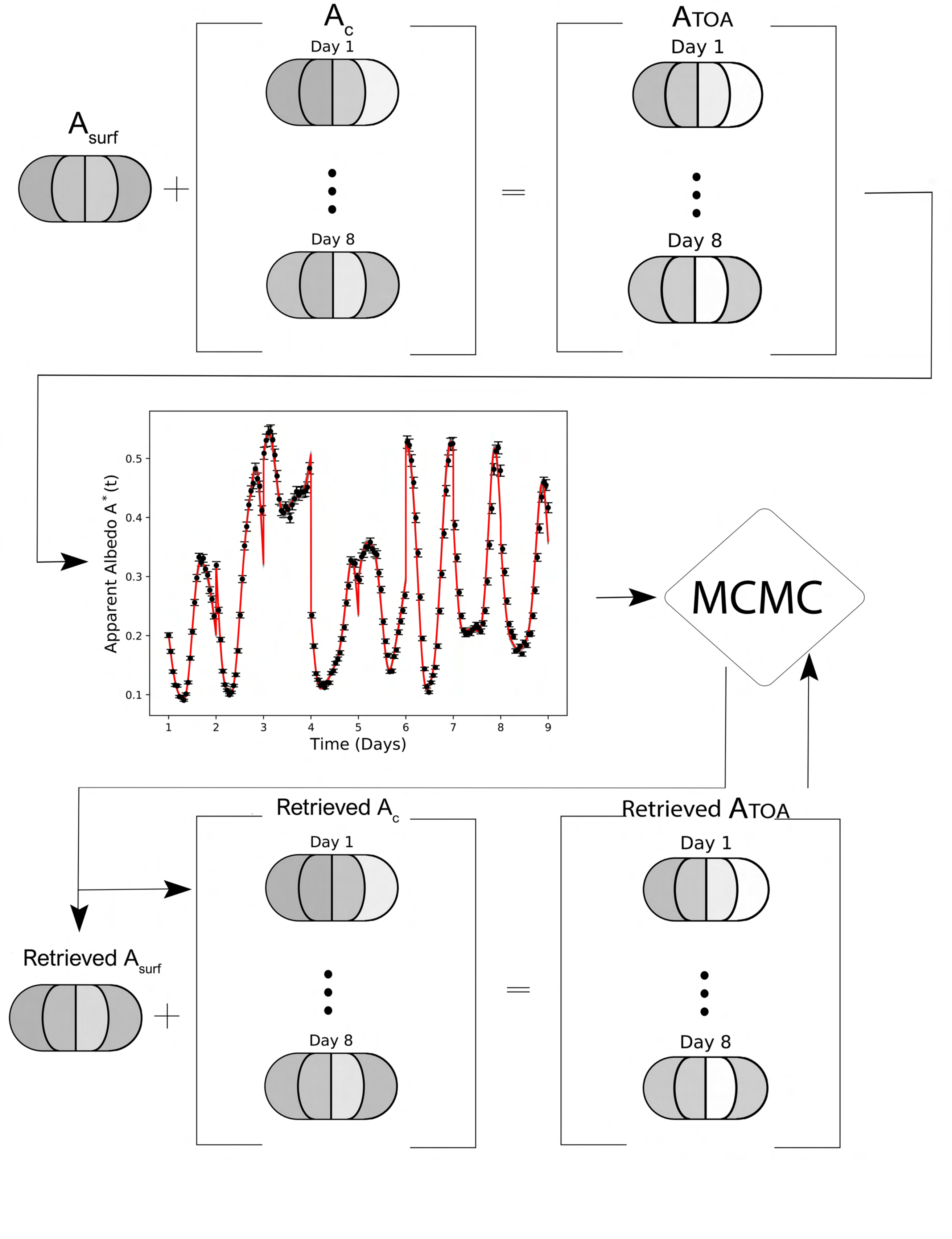}
    \caption{Schematic description of the different steps of our numerical experiment, from producing the synthetic light curves to the MCMC retrieval of the surface albedo and time-varying clouds maps, $\Delta A$. We only show days 1 and 8 in this illustration but the surface map is fitted using all eight days of data. A$_{\rm TOA}$ stands for the top-of-atmosphere albedo map.}
    \label{moneyplot}
\end{figure*}

To generate synthetic data we choose a number of longitudinal slices for our map, then randomly generate a surface albedo map and daily cloud albedo maps, $\rm A_c$. Using the appropriate $\rm A_{c,i}$ and Eq. (\ref{effAlb}), we compute the top-of-atmosphere albedo of each slice for that day. Our default range for the surface albedo is between 0 and 0.5, and between 0 and 1 for $\rm A_c$. Both sets of parameters are generated using a uniform distribution between the mentioned bounds. This ensures that clouds randomly change once a day. We then use Eq. (\ref{forward_fin}) to compute a light curve and add random Gaussian noise with a standard deviation equal to 2\%  of the mean of the light curve to simulate photometric uncertainties. These are optimistic photometric uncertainties for next-generation space telescopes (\citealt{cowan_alien_2009, fujii_colors_2010}). An example synthetic light curve is shown in the middle of Fig. \ref{moneyplot}, along with the best-fit model from the Markov Chain Monte Carlo described below.

\section{Removing Clouds from Synthetic Data} \label{Section : remov synth data}
Fig. \ref{moneyplot} summarises the process of generating a synthetic light curve and then retrieving surface and cloud maps.

\subsection{Inverse Problem} \label{Inverse PB}
We wish to retrieve the surface albedo map and time-variable cloud map of an unresolved planet based on its quasi-periodic rotational light curve.
We tackle the inverse problem by repeatedly computing light curves within a Markov Chain Monte-Carlo  \citep[specifically \texttt{emcee};][]{Foreman_Mackey_2013}.  

The natural logarithm of the likelihood of an observed lightcurve $a^*(t)$ given a model $A^*(t)$ is
\begin{equation} \label{likelihood)} \small
    \ln{\left(p(a^*|t,A^*,\sigma)\right)} = -\frac{1}{2} \Bigg[\sum_{i = 1}^{n}\frac{(a_i^*-A_i^*)^2}{\sigma_i^2} + \ln{(2 \pi \sigma_i^2)}\Bigg],
\end{equation}
where $a_i^*$ is a datum, $\sigma_i$ its uncertainty, and $A_i^*$ is the prediction of the forward model. We adopt uniform priors on $\rm A_{{\rm s},i}$ and $\ln(\rm A_{c,i,j})$. The log-uniform prior on cloud albedo means that clouds are kept to the minimal value allowed by the data, thus breaking the surface--cloud degeneracy manifest in Eq.\ \ref{effAlb} (attempts with uniform prior on $\rm A_c$ never converge). Since all of our priors are uniform, the posterior on the parameters is proportional to the likelihood.   We constrain the values of $\ln{(\rm A_c)}$ to be between -6 and 0 and $\rm A_s$ to be between 0 and 0.5. \par 

The parameter space has dimensions of $n_{\rm sl}\times (n_{\rm day}+1)$, where $n_{\rm sl}$ is the number of slices in the map and $n_{\rm day}$ is the number of days spanned by the data. The number of slices in the retrieval maps needs not equal the resolution of the map that produced the lightcurve.  We do not fit for the planet's rotational period; previous work has demonstrated that it can readily be measured with time-resolved photometry (\citealt{2008ApJ...676.1319P,2009ApJ...700.1428O}). 

In order to ensure that we have fewer fitted parameters than data, we derive an upper limit on the number of slices one should use for a retrieval, as a function of $n_{\rm day}$ and $n_{\rm data}$, the number of data per day. This upper limit is given by 
\begin{equation} \label{nslices max}
    n_{\rm sl}^{\rm max}  = \frac{n_{ \rm data} \times n_{ \rm day}}{n_{ \rm day}+1} .
\end{equation}
For the EPIC data that we fit in Section \ref{EPIC}, $n_{\rm data}$ ranges between 13 and 22. An 8 day simulation with 22 data points per day yields a $n_{\rm sl}^{\rm max}=19$, while for 13 data points per day, $n_{\rm sl}^{\rm max}=11$. In practice, however, we are constrained to far fewer slices given the lossy nature of the map-to-lightcurve transformation.

Our experiment is run at full phase to mimic the viewing geometry of the Deep Space Climate Observatory, so we always have a convolution kernel spanning half of the planet. In order to achieve Nyquist sampling of the convolution kernel, we need 4 slices, more than that will lead to over-fitting (\citealt{cowan_alien_2009,cowan_rotational_2011}). Therefore, our fiducial model uses four slices.

For this work, the walkers used for the MCMC retrievals are initialised in a Gaussian ball around best-fit values found using a gradient descent algorithm.  
We use 200 walkers for each retrieval. To check if the MCMC has converged, we iteratively check by eye the trace of the walkers. We choose to run 20 000 steps and to burn the first 15 000, as we find that it provides a good balance between convergence and computational time. 

\subsection{How many planetary rotations are required to remove clouds?}\label{day_remove_clouds}
We now determine the number of planetary rotations needed to retrieve an accurate surface albedo map and strip the effect of clouds from a synthetic light curve. The EPIC observations have 13 to 22 data per rotation, so we generate synthetic light curves with 22 data per rotation, the most favourable scenario for EPIC data. Since the number of cloud parameters is proportional to the number of planetary rotations considered, choosing a number of days is equivalent to choosing the number of fitted parameters. With a four slice map, we generated synthetic lightcurves spanning 1, 2, 4, 8, or 16 rotations. Each simulation was repeated thirty times, using a different map and light curve each time. We then define the surface albedo accuracy  as $A_{ \rm surf,fit}-A_{ \rm surf, true}$, and the surface albedo precision as the mean of the estimated uncertainties.

\begin{figure}
    \centering
    \includegraphics[width=0.48\textwidth]{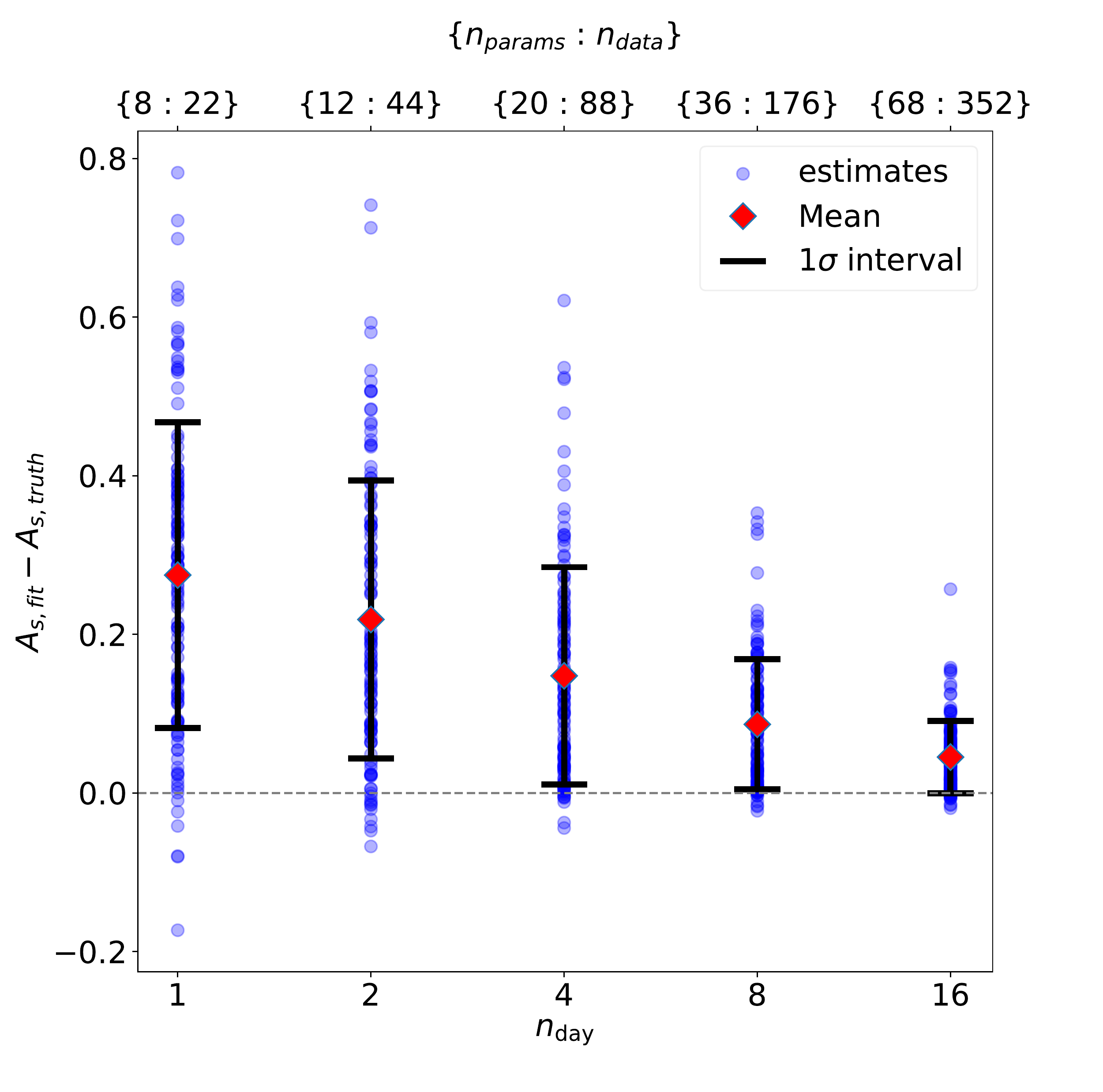}
    \caption{Accuracy and precision of surface albedo as a function of the number of days---or rotations of the planet---used in the retrieval. For each $\rm n_{day}$ we produce 30 lightcurves. Each blue dot represents a single slice of a single simulation, and the red diamond is the overall mean and hence an estimate of bias (closer to zero is better). The standard deviation is shown as error bars and is indicative of accuracy (smaller is better). We use maps with 4 longitudinal slices to generate the synthetic light curves and to fit them. Retrieving a surface map from a single rotation is hopeless because the surface and clouds maps are entirely degenerate, but with eight days of photometry the retrieved surface albedo becomes less biased and more accurate.}
    \label{acc prec diff nday}
\end{figure}

As shown in Fig. \ref{acc prec diff nday}, the bias on the surface albedo is reduced when fitting more days. The number of fitted parameters and the number of data are listed at the top of Fig. \ref{acc prec diff nday}---both increase when considering more days but the fit becomes more constrained. When using only one planetary rotation the retrieved surface albedo map is very poorly constrained because the surface and cloud maps are entirely degenerate; retrievals performed on two and four rotations only fare a bit better. The surface albedo is always over-estimated indicating that cloud removal is imperfect even with 16 rotations of data. We adopt 8 days for our fiducial model as a compromise between retrieval accuracy and the increasing programmatic challenge of long continuous observations of directly-imaged exoplanets. \par
Log-uniform priors on the cloud albedo bias our cloud estimates low. This leads to an overestimation of the surface albedo that can only be ameliorated with longer continuous observations. Indeed, the longer the observations, the better we break the surface-cloud degeneracy. This trade-off between clouds-surface degeneracy and biased clouds is necessary since linear (unbiased) priors are unable to reach converged states. Therefore, one needs to use continuous observations on the timescale of days to hope to break the bias introduced by our model. \par

We also tested whether increasing the data points per day reduced the bias in retrieved surface albedo. Unsurprisingly, we find a slight improvement of the retrievals when using twice as many points per day. In practice, however, the time resolution of data will likely be anti-correlated with their precision so we do not further explore this parameter. 
\subsection{Results with 4 longitudinal slices and 8 days worth of synthetic data} \label{res_sim_data}
To assess the fidelity of our retrievals, we use the Z-score metric,
\begin{equation} \label{Zscore} 
     Z_{\rm i} = \frac{A_{{\rm fit},i}-A_{{\rm true},i}}{\sigma_i} ,
 \end{equation}
where $\sigma_i$ is the estimated uncertainty on the albedo  of slice $i$ (half the difference between the 84$^{\rm th}$ and the 16$^{\rm th}$ percentiles of the marginalised posterior). Table \ref{tab:summary} summarises the accuracy and precision of retrieved maps.

\begin{table*}
 \caption{Fidelity of retrieved map parameters. 
 $\Delta$ Range: minima and maxima of $ \Delta = { \rm fit}-{\rm true}$. Bias: mean of $\Delta$. Accuracy: standard deviation of $\Delta$. Precision: mean of the estimated uncertainty on a given quantity; Mean Z-score: how many $\sigma$ away from the truth the estimates are; Standard deviation of Z-scores: the fidelity of the uncertainty estimates. For bias, accuracy, precision and mean z-scores, smaller values are better. For the standard deviation of z-scores, values close to one reflect honest uncertainty estimates. The second to last row computes the metrics for the retrieved TOA albedo with respect to the true surface albedo, $A_{\rm TOA}-A_{\rm s,truth}$. Notably, the top-of-atmosphere albedo, while easier to retrieve, severely over-estimates the surface albedo. The last row computes the metrics using the minimal retrieved TOA albedo as the surface albedo. In this scenario of 2\% photometric uncertainties, this approach yields similar results than using the fitted surface albedo.}
 \label{tab:summary}
 \begin{tabular}{|l|c|c|c|c|c|r|}
    \hline
         & $\Delta$ Range& Bias & Accuracy &Precision& Mean Z-score & Std of Z-score \\
         \hline
         $A_{\rm TOA}-A_{\rm TOA, truth}$  & $[-0.043 : 0.049]$ & 7.7 $\times$ 10$^{-4}$ &  1.3 $\times$ 10$^{-2}$ & 2.0 $\times$ 10$^{-2}$ & -0.054 & 0.72\\
         \hline
         $ A_c- A_{\rm c, truth}$  &$[-0.37:0.033]$ & -8.7 $\times$ 10$^{-2}$ & 8.3 $\times$ 10$^{-3}$ &1.6 $\times$ 10$^{-2}$ & -6.03 & 7.11\\
         \hline
         $A_{\rm s}-A_{\rm s,truth}$ & $[-0.022 : 0.35]$ & 8.6 $\times$ 10$^{-2}$&  8.2 $\times$ 10$^{-2}$ & 1.2 $\times$ 10$^{-2}$ & 7.21& 7.44 \\ 
         \hline
         $A_{\rm TOA}- A_{\rm s,truth}$ &$[-0.013:0.91]$ & $4.4\times 10^{-1}$ & $2.6 \times10^{-1}$ &$2.0 \times 10^{-2}$ & 22.01& 12.94\\
         \hline
         A$_{\rm TOA,min}$-A$_{\rm s, truth}$ & $[-0.013:0.35]$ & 9.2 $\times$ 10$^{-2}$ & 8.2 $\times$ 10$^{-2}$ & 2.0 $\times$ 10$^{-2}$ & 5.65 & 5.44 \\
         \hline 
 \end{tabular}
\end{table*}

Our retrieval yields a top-of-atmosphere albedo maps for each day of observation, since the cloud cover changes from one day to another. The top-of-atmosphere maps, $A_{\rm TOA}$, show little bias and a good accuracy. We slightly overestimate the uncertainties as the standard deviation of the Z-score is only around 0.72.

Our estimates of the clouds, $\rm A_c$, are less good than for the top-of-atmosphere albedo. This makes intuitive sense because the clouds only indirectly impact the lightcurve and each slice of a cloud map only affects a single day of data. The cloud pattern is utterly different from one day to the next, so the cloud cover is hard to constrain. The bias is negative for the clouds estimates, which means that our model slightly underestimates the albedo contribution of clouds.

We can now analyse the surface maps, arguably the most interesting features retrieved by our model. The retrieved surface albedo maps, $A_s$, show a small positive bias due to the imperfect cloud removal but have a good accuracy of 0.082 (Table \ref{tab:summary}). Therefore, our retrieved maps are in good agreement with the maps used to create the synthetic data. However, the standard deviation of the Z-score are greater than the expected values of 0 and 1 respectively. This shows that we over-estimate the surface albedo and under-estimate its uncertainty.

The second to last row of Table \ref{tab:summary} compares the retrieved TOA albedo map to the true surface albedo. This is the status quo approach of optimistically treating the TOA map as a surface map.  The retrieved mean Z-score shows that our TOA albedo map is 22$\sigma$ too high, on average; this performance is about 7$\times$ worse than our cloud removal scheme. Indeed, as the clouds are more reflective than the surface, interpreting the TOA albedo as the surface albedo lead to a dramatic overestimation of the surface albedo.

The last row of Table \ref{tab:summary} interprets the minimal TOA albedo of a slice as the surface albedo of that slice. This approach yields comparably good estimates of the surface albedo maps. 
However, in experiments with greater uncertainties (4\% rather than 2\%), the fidelity of the minimum TOA albedo approach deteriorates much more than actually fitting the surface concurrently with the clouds.

\section{Removing Clouds from the EPIC Data} \label{EPIC}
In this section, we apply our cloud removal scheme to observations of Earth in an attempt to produce a cloudless map of its surface. Specifically, we use data from the Deep Space Climate Observatory's Earth Polychromatic Imaging Camera  \citep[DSCOVR EPIC;][]{2018BAMS...99.1829M}. This satellite is situated at the Sun--Earth system's first Lagrange point, and has spent years observing the day side of Earth. In order to mimic observations of a directly-imaged exoplanet, each EPIC image is averaged to a single apparent albedo datum \citep{jiang_using_2018} \citep[for a review of Earth-as-an-exoplanet experiments, see][]{2018arXiv180404138R}.

\subsection{EPIC data} \label{EPIC DATA}
EPIC acquires images of Earth at 10 different wavelengths every 68 to 110 minutes. Therefore, it takes between 13 and 22 images per day \citep{jiang_using_2018}. The light curves acquired by this instrument for all ten wavelengths were provided by J.H. Jiang. More than two years of data were provided, starting on June 2015 at 00:00:00 UTC. For the purpose of this study, we only use the 551nm and the 779.5 nm EPIC channel, since cloud coverage is respectively the most and the least impactful at these wavelengths according to \cite{jiang_using_2018}. We adopt the same uncertainties for the EPIC data as for the synthetic lightcurves: Gaussian noise with a standard deviation equal to 2\% of the mean of the lightcurve considered.

\begin{figure*}
    \centering
    \includegraphics[scale = 0.65]{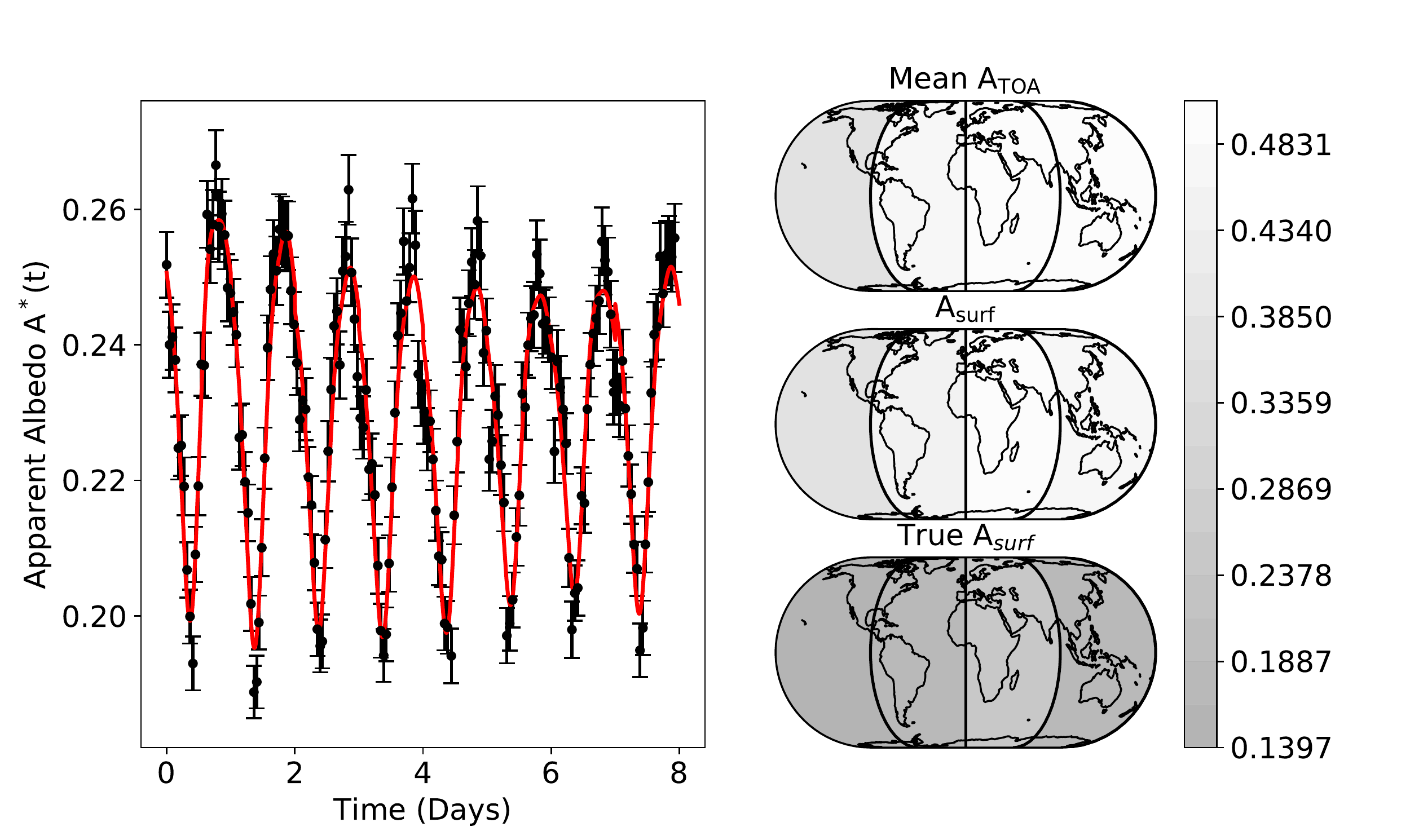}
    \caption{\emph{Left:} EPIC 779.5 nm lightcurve starting on June 12$^{\rm th}$ 2017. Black dots with uncertainties represent the EPIC data. We adopt photometric error bars equal to 2\% of the mean apparent albedo. The red line is the MCMC fit to the data. \emph{Right:} from top to bottom, the mean retrieved top-of-atmosphere albedo map, the retrieved surface albedo, and the true surface albedo computed using the MODIS dataset. Mean top-of-atmosphere albedo and surface maps only differ by a few percent, on the order of the standard deviation in daily retrieved top-of-atmosphere maps. Our cloud removal efforts are stymied by the modest cloud variability of Earth on 90$^\circ$ length scales.}
    \label{lastfig}
\end{figure*}

We fitted 8 days of EPIC photometry starting on day 730, i.e., on the 12$^{ \rm th}$ of June 2017 (see Fig. \ref{lastfig}).  We use this 8-day stretch as it is the time-series with the highest-cadence observations in the entire data set.

\subsection{Results on EPIC data} \label{EPIC RESULTS subsec}

 We performed retrievals on the EPIC data in the same manner as the retrievals on synthetic lightcurves: \texttt{emcee} with 200 walkers and 20 000 steps, burning in for the first 15 000 steps, to retrieve 4-slice maps of surface and cloud albedo. Fig. \ref{lastfig} shows the fitted EPIC 779.5 nm light curve as well as the surface albedo map and the eight day average top-of-atmosphere map. The bottom panel shows the ground truth we hoped to recover, computed using MODIS data\footnote{\url{https://modis.gsfc.nasa.gov/data/}} following \cite{cowan_alien_2009}.

Our retrieved surface albedo map is a poor approximation of the surface features of Earth. In fact the retrieved surface albedo is only 1.4\% lower than the TOA albedo, on average. The small difference between top-of-atmosphere and surface maps is unsurprising given the lack of variability in TOA albedo from one day to another. Indeed, the standard deviation of the daily top-of-atmosphere maps on all slices is less than 1\%. The 551 nm channel performs even more poorly. The top-of-atmosphere albedo varies half as much in this channel as at 779.5 nm so essentially all of the albedo is attributed to a static surface rather than clouds. Our approach is limited to only remove clouds insofar as they vary. \par

\section{Discussion} \label{Discuss}

Despite encouraging performances on synthetic lightcurves, our cloud removal scheme struggles on the real clouds of EPIC data.  The EPIC lightcurves change by only 1\% from one rotation to another so our method was only able to reduce the impact of clouds at that level. The small change in $A_{\rm TOA}$ was unexpected because the characteristic timescale for clouds to change on Earth is a few days \citep[][]{RevModPhys.56.365}, shorter than the eight day observing campaigns we fitted. 

The poor performance of our cloud removal scheme for Earth is likely due to the small spatial scale of weather patterns combined with unfavourable viewing geometry. We performed our experiment at full phase in order to match the EPIC viewing geometry, observing Earth from L$_1$. We average the local variability of clouds over a whole hemisphere, so it evens out to a small global variability in EPIC data. 
Observing Earth at larger phase angles would reduce this smoothing of local cloud variability and allow us to better remove clouds from the retrieved surface maps. Moreover, exoplanets cannot be directly imaged near full phase since the coronagraph/starshade gets in the way. 

Direct imaging is most likely to see planets near quadrature (quarter phase). In principle, this should facilitate the mapping and removal of clouds: at quarter phases the convolution kernel is narrower than the one we used at full phase, so the phase-dependent maximal number of slices would be closer to 8 than to 4  \citep{cowan_alien_2009,cowan_rotational_2011}. Moreover, at quarter phases,  the scattering phase function should be close to Lambertian, unaffected by the forward and back-scattering peaks that impact observations at crescent and full phase (\citealt{burrows2010giant,robinson_detecting_2010}).\par 
We performed simulations in the same fashion as in Section \ref{Section : remov synth data} but at quarter phase; we again assumed edge-on orbit and zero obliquity. We tested the impact of the numbers of days on the retrieved quantities and unsurprisingly reach the identical conclusions. Moreover, we tested the effect of using 8 slices instead of 4 in the retrievals. Our results show a bigger range (in the sense of the 1st column of Table \ref{tab:summary}) of the retrieved parameters. This intuitively makes sense because the spatial resolution is increased while keeping the amount of available information constant. We find very similar bias, accuracy and precision. In particular, the positive bias of the surface albedo is still present, a manifestation of the log-uniform priors on the clouds (see Section \ref{day_remove_clouds}). However, we find that the mean Z-score and associated standard deviations are better constrained than in the full phase case. This confirms for surface albedo that the quarter phase geometry is more favourable to mapping. \par

Our choice of 4 slices for our retrievals was motivated by the width of the convolution kernel at full phase. Surface features on Earth happen to have a characteristic scale of $\sim$10$^4$ km \citep{farr_exocartographer_2018}, one quarter the circumference of the planet. Nonetheless, there is no reason to believe that surface and cloud features on an exoplanet always span a quarter of the globe. In order to quantify the impact of intrinsic vs.\ retrieved length scales, we generated thirty light curves with two and eight-slice planets, each spanning eight days, and we tried to retrieve the surface parameters using models with four slices. Our results are summarised in Table \ref{tab:summary2}.

\begin{table*}
 \caption{Fidelity of fitted parameters for a retrieval with 4-slice maps.  The true input maps used to generate the lightcurves had resolutions of 2, 4 or 8 slices (the middle two rows are duplicated from Table 1). There is no problem in using too high a spatial resolution when performing a retrieval, but using too low a resolution leads to a biased surface map.}
 \label{tab:summary2}
 \begin{tabular}{|l|c|c|c|c|c|c|c|r|}
    \hline
         & N$_{\rm sl, truth}$& $\Delta$ Range& Bias & Accuracy &Precision& Mean Z-score & Std of Z-score \\
         \hline
         $A_{\rm TOA}-A_{\rm TOA, truth}$ &2 & $[ -0.016:-0.016 ]$ &9.8 $\times 10^{-4}$  &5.2$\times 10^{-3}$  &1.5 $\times 10^{-2}$  &0.049& 0.39 \\
         \hline
         $A_s - A_{\rm s, truth}$ & 2 &$[-0,32:0.61]$ & 1.1 $\times$ 10$^{-1}$ & 1.7 $\times$ 10$^{-1}$ &9.8 $\times$ 10$^{-3}$ & 16.1 & 26.95\\
         \hline
         $A_{\rm TOA}-A_{\rm TOA, truth}$  &4&  $[-0.043 : 0.049]$ & 7.7 $\times$ 10$^{-4}$ &  1.3 $\times$ 10$^{-2}$ & 2.0 $\times$ 10$^{-2}$ & -0.054 & 0.72\\
         \hline
         $A_{\rm s}-A_{\rm s,truth}$ & 4 & $[-0.022 : 0.35]$ & 8.6 $\times$ 10$^{-2}$&  8.2 $\times$ 10$^{-2}$ & 1.2 $\times$ 10$^{-2}$ & 7.21& 7.44 \\
         \hline
         $A_{\rm TOA}-A_{\rm TOA, truth}$ & 8 &$[-0.21:0.21]$ &-6.5 $\times 10^{-3}$  &6.8 $\times 10^{-2}$ &2.5 $\times 10^{-2}$ &-0.46  &3.14 \\
         \hline
         $A_s - A_{\rm s, truth}$ & 8 &$[-0.16:0.43]$ & 1.6 $\times$ 10$^{-1}$ & 1.2 $\times$ 10$^{-1}$ &1.7 $\times$ 10$^{-2}$ & 11.01 & 9.63\\
         \hline
 \end{tabular}
\end{table*}

There is no problem in using too high a spatial resolution when performing a retrieval, but using too low a resolution leads to a biased surface map.  When retrieving with a 2-slice map, we find that the retrieved surface maps show slightly less bias, better accuracy, precision and Z-score than the 4-slice planet, but this is expected from the central limit theorem: it is always easier to estimate averaged quantities. When retrieving an eight-slice planet with a four-slice model, however, the surface albedo is biased high because cloud removal is less effective. 
This may explain the disappointing results of our cloud removal on the EPIC data: cloud patterns on Earth are usually less than 90$^\circ$ across.

\section{Conclusion}\label{conclu}
We developed a framework to map the clouds and surface of unresolved exoplanets using reflected light photometry. We tested our model on synthetic 8-day light curves of a hypothetical planet with randomly varying clouds, and on eight days of EPIC photometry of Earth. 
When applied to synthetic data, our cloud removal scheme is capable of recovering surface albedo maps with little bias, good precision and accuracy, but slightly underestimated uncertainties. 

Our retrievals on synthetic data provide acceptable estimates of surface albedo.  Indeed, when photometric uncertainties are negligible, simply interpreting the lowest measured top-of-atmosphere albedo at each location provides a good estimate of the surface albedo, but such quick-and-dirty inferences should be treated with caution.  Regardless of how they are made, such low-resolution surface maps could help identify exoplanets with continents and oceans, and hence prospects for long-term habitability \citep[][]{abbot_indication_2012}. 

Tests on real lightcurves of Earth and synthetic lightcurves produced with higher resolution maps result in poor cloud removal. We therefore conclude that exoplanets with large, changing cloud structures observed near quadrature phases would be ideal candidates for cloud removal.

 Our cloud modelling and removal scheme could in principle be combined with multi wavelength light curves, which have previously been show to help identify clouds in single-day light curves (e.g., \citealt{cowan_alien_2009, jiang_using_2018}). For suitable planets and viewing geometries, cloud removal yields more information than the top-of-atmosphere albedo map: in addition to a surface map, our approach provides daily cloud maps and an estimation of the mean cloud albedo of the planet.
 
\section*{Acknowledgements}

We thank Jonathan Jiang for sharing disk-integrated EPIC photometry and the McGill Exoplanet Characterisation Alliance (MEChA) graduate students for their help during this study. We acknowledge the camaraderie and support of the McGill Space Institute (MSI) and l'Institut de recherche sur les  exoplanètes (iREx). This work made use of the McGill Physics computing cluster. We also thank both anonymous referees for constructive comments that significantly improved the first version of this manuscript.

\section*{Data Availability Statement}
The data underlying this article will be shared on reasonable request to the corresponding author.

\section*{Code Availability Statement}
The code is publicly available on GitHub, at this repository: \url{https://github.com/LTeinturier/Cloud-Killer-}




\bibliographystyle{mnras}
\bibliography{Teinturier_al} 




\appendix
\section{Derivation of the Forward Model at different phases } \label{appendix}
Here, we derive the computation of the forward model for an arbitrary phase angle $\alpha$, not necessarily full phase. \par
One can rewrite the convolution kernel as 
\begin{equation}
    \begin{split}
     &K(\Omega,t) = \\
     &\frac{1}{\pi} \max[\sin{(\theta)}\sin{(\theta_s)}\cos{(\phi-\phi_s)}+\cos{(\theta)}\cos{(\theta_s)},0] \\
     & \times \max[\sin{(\theta)}\sin{(\theta_o)}\cos{(\phi-\phi_0)}+\cos{(\theta)}\cos{(\theta_o)},0]
    \end{split}
\end{equation}
We continue with the simplifying assumption of zero obliquity and an edge-on orbit:  $\theta_s$ = $\theta_o$ = $\frac{\pi}{2}$ and $\phi_s$ = $\phi_o$ + $\alpha$. This yields, for $\theta \in [0: \pi]$ and $\phi \in [\phi_o+\alpha-\frac{\pi}{2}: \phi_o+\frac{\pi}{2}]$
\begin{equation}
    \begin{split}
     K(\Omega,t) &= \frac{1}{\pi}\max[\sin{(\theta)}\cos{(\phi-\phi_o),0]} \\
     &\times \max[\sin{(\theta)}\cos{(\phi-\phi_o-\alpha)},0] \\ 
    & = \frac{1}{\pi}\big(\sin^2{(\theta)}\cos{(\phi-\phi_o)}\cos{(\phi-\phi_o-\alpha)}\big)
    \end{split}
\end{equation}
Using the trigonometry identity $\cos{p}\cos{q}=\frac{1}{2}\cos{(p+q)}\cos{(p-q)}$, we derive the convolution kernel, depending on the phase angle
\begin{equation}
    K(\Omega,t) = \frac{1}{\pi}\sin^2{(\theta)}\times \frac{1}{2}\Big(\cos{(\alpha)}+\cos{(2\phi-2\phi_o-\alpha)}\Big)
\end{equation}
We can now input this phase angle-dependant convolution kernel into Eq. \ref{forwardmodel} to get the apparent albedo, using $d \Omega = \sin(\theta)d\theta d\phi $ : 
\begin{equation}
    A^*(t) = \frac{\oint K(\Omega,t)A_{\rm TOA}(\Omega)d\Omega}{\int_0^{\pi}\sin^3{\theta}d\theta\int_{\phi_o+\alpha-\frac{\pi}{2}}^{\phi_o+\frac{\pi}{2}}\frac{1}{2}\big(\cos{(\alpha)}+\cos{(2\phi-2\phi_o-\alpha)}\big)d\phi}
\end{equation}
In the denominator. $\int_0^\pi \sin^3{\theta}d\theta$ = $\frac{4}{3}$, and 
\begin{equation}
    \begin{split}
    & \int_{\phi_o+\alpha-\frac{\pi}{2}}^{\phi_o+\frac{\pi}{2}} \frac{1}{2}\Big(\cos(\alpha)+\cos(2\phi-2\phi_o-\alpha)\Big)d\phi \\
    & = \frac{1}{2}(\pi-\alpha)\cos{(\alpha)} \\
    & +\frac{1}{4}\Big(\sin{(2\phi_o+\pi-2\phi_o-\alpha)}-\sin({2\phi_o+2\alpha-\pi-2\phi_o-\alpha)}\Big) \\
    & = \frac{1}{2}(\pi-\alpha)\cos(\alpha)+ \frac{1}{2}\sin{(\pi-\alpha}).
    \end{split}
\end{equation}
Now, we discretize the numerator on a slice [$\phi_i:\phi_{i+1}$], taking into account the previous computations, and integrating the numerator over the co-latitude $\theta$ : 
\begin{equation}
    \begin{split}
        A_i^*(t) &= A_i^{\rm TOA}\frac{\int_{\phi_i}^{\phi_{i+1}}\big(\cos(\alpha)+\cos(2\phi-2\phi_o-\alpha)\big)d\phi}{(\pi-\alpha)\cos(\alpha)+\sin(\pi-\alpha)} \\
        & = A_i^{\rm TOA}\frac{\Big[\phi \cos(\alpha) + \frac{1}{2}\sin(2\phi-2\phi_o-\alpha)\Big]_{\phi_i}^{\phi_{i+1}}}{(\pi-\alpha)\cos(\alpha)+\sin(\pi-\alpha)}.
    \end{split}
\end{equation}
We get the apparent albedo of the planet by summing on all visible and illuminated slices $m$, 
\begin{equation}
    A^*(t) = \sum_{i=1}^{m}\frac{A_i^{\rm TOA}\Big[\phi \cos(\alpha)+\frac{1}{2}\sin(2\phi-2\phi_o-\alpha)\Big]_{\max(T_W,\phi_i)}^{\min(T_E,\phi_{i+1})}}{(\pi-\alpha)\cos(\alpha)+\sin(\pi-\alpha)}
\end{equation}
In the case of quarter phase ($\alpha = \frac{\pi}{2}$), the apparent albedo is 
\begin{equation}
    A^*(t) =\frac{1}{2} \sum_{i=1}^{m}\Big[\sin(2\phi-2\phi_o-\frac{\pi}{2})\Big]_{\max(T_W,\phi_i)}^{\min(T_E,\phi_{i+1})}.
\end{equation}
At full phase ($\alpha=0$), we find Eq. \ref{forward_fin}.


\bsp	
\label{lastpage}
\end{document}